\documentclass[onecolumn,floatfix,showpacs,aps,nofootinbib, showkeys]{revtex4}

\usepackage[dvips]{epsfig}
\usepackage[english]{babel}
\usepackage{bbm}
\usepackage{array}
\usepackage{amsmath}
\usepackage{multirow}
\usepackage{amsfonts}
\usepackage{amssymb}
\usepackage{hyperref}
\usepackage{leading}
\usepackage[dvipsnames]{xcolor}
 
\hypersetup{colorlinks=true,linkbordercolor=Blue,linkcolor=Blue, citecolor=Blue}
\usepackage{subeqnarray}
\usepackage{setspace}
\usepackage{indentfirst} %primeiro paragrafo com espaço
\usepackage{gensymb}%simbolo º

\AtBeginDocument{
    \hypersetup{colorlinks,urlcolor=Blue}
}

\begin{document}

\title{Flag-dipole spinors: On the dual structure derivation and $\mathcal{C}$, $\mathcal{P}$ and $\mathcal{T}$ symmetries}

\author{R. J. Bueno Rogerio$^{1}$} \email{rodolforogerio@unifei.edu.br}
\author{A. R. Aguirre$^{1}$} \email{alexis.roaaguirre@unifei.edu.br}
\author{C. H. Coronado Villalobos$^{2}$} \email{ccoronado@autonoma.edu.pe}
\affiliation{$^{1}$Instituto de F\'isica e Qu\'imica, Universidade Federal de Itajub\'a - IFQ/UNIFEI, \\
Av. BPS 1303, CEP 37500-903, Itajub\'a - MG, Brazil.}
\affiliation{$^{2}$Universidad Aut\'onoma del Per\'u, Facultad de Ingenier\'ia y Arquitectura, Escuela Profesional de Ingenier\'ia de Sistemas\\
Panamericana Sur Km 16.3, Villa el Salvador, Lima-Per\'u.}

%\date{\today}

\begin{abstract}
\noindent{\textbf{Abstract.}}
In this manuscript we report the flag-dipole spinors dual structure direct definition and analyze the properties behind the corresponding operator which generates such structure. This particular construction may be interesting for cosmological, phenomenological and mathematical physics applications. In addition, we analyse the behaviour of the flag-dipole spinors under action of discrete symmetries, facing an \emph{unconventional} property encoded on $(\mathcal{CPT})^2$.

\end{abstract}


\maketitle

\section{Introduction}\label{intro}

Spinors play an important role in several areas of Quantum Field Theory. Such mathematical objects must be understood as an irreducible representation of the Lorentz group $SO_{+}(1,3)$ \cite{Wigner1,Wigner2,ryder}, which carry an extensive information about the space-time in which they are defined. All relevant physical information associated with the spinors is encoded in its bilinear forms. In fact, some years ago an spinor classification based on bilinear covariants and multivectors of observables was developed by Lounesto \cite{lounestolivro}. 

Such classification sheds light on the existence of new classes of spinors. In particular, it revealed  the so-called flag-dipole spinors, which reside between the Weyl, Majorana and Dirac spinors. It is of common knowledge that the relativistic description of the electron allow one to define the following set of bilinear forms: the invariant length $\sigma=\bar{\psi}\psi$, the pseudo scalar amount $\omega=\bar{\psi}\gamma^{5}\psi$, the current density defined as $\boldsymbol{J}=\bar{\psi}\gamma^{\mu}\psi\gamma_{\mu}$, the spin projection in the momentum direction $\boldsymbol{K}=\bar{\psi}\gamma^{\mu}\gamma^{5}\psi\gamma_{\mu}$, and the momentum electromagnetic density $\boldsymbol{S}=\bar{\psi}i\gamma^{\mu\nu}\psi\gamma_{\mu}\wedge\gamma_{\nu}$, where we have defined $\bar{\psi}=\psi^{\dag}\gamma_0$ and  $\gamma$ stands for the Dirac matrices \cite{lounestolivro,crawford1}. A more comprehensive description for bilinear forms can be found at \cite{beyondlounesto}. The 16 aforementioned bilinear forms are restricted to obey an algebraic quadratic relation known as Fierz-Pauli-Kofink identities \cite{lounestolivro}. 

The Lounesto's classification can be divided into two sectors, one embracing single-helicity spinors (classes 1, 2 and 3) and the other dual-helicity spinors (classes 4 and 5) \cite{interplay,rodolfoconstraints,rodolfosubliminal}. The first three classes of the Lounesto's classification describe the Dirac spinors. The fourth class consists of flag-dipole spinors with a flag $\boldsymbol{S}$ on a dipole of two poles $\boldsymbol{J}$ and $\boldsymbol{K}$. The fifth class (Majorana spinors) consists of flag-pole spinors with a flag $\boldsymbol{S}$ on a pole $\boldsymbol{J}$, and the sixth class (Weyl spinors) consists of dipole spinors with two poles $\boldsymbol{J}$ and $\boldsymbol{K}$ \cite{lounestolivro}. At this point, it is worth mentioning that the properties of the flag-dipole spinors had not been yet defined properly, only slightly explored in some very specific scenarios \cite{da2010elko,esk,roldaomeert,roldaonewspinor,fabbrifr,vignolofr,ferrariflagdipole,roldaorogerioflagdipole}. Therefore, it is one of the purposes of this communication, to report and to describe some of the properties related to such spinor.

Quite recently, interesting new insights were brought to scene after the Elko's theoretical discovery \cite{mdobook}. Proposed in its first formulation in 2004 \cite{jcap}, the spin-${1/2}$ fermionic field endowed with mass dimension one, constructed upon a complete set of eigenspinors of the charge conjugation operator and, consequently, due to its restricted interactions with the Standard Model particles, is believed to be a strong candidate to describe dark matter \cite{mdobook}. 

The features carried throughout mass dimension one theory, opened windows to a physical content known as \emph{Beyond the Standard Model Theory}, trying to answer many questions that seems to be incomplete. Since the mass dimension one theory is still being constructed \cite{mdobook,tipo4epjc}, we believe that a broad and quite interesting content is hidden beyond the mass dimension one fermions. So far, in the literature, we have two theoretical examples of mass dimension one fermions: the Elko and the flag-dipole spinors. Both spinors carry quite  peculiar and particular features. Regarding these particularities, we are able to list: a new dual structure, the dynamics, and their unconventional (meanwhile \emph{expected}) behaviour under $\mathcal{C}$, $\mathcal{P}$ and $\mathcal{T}$ discrete symmetries. 

The focus of the present manuscript is to exhibit in detail an \emph{Ab Initio} construction of the flag-dipole spinor dual structure. Interesting enough, the contrasting dual structure carry an involved operator, which is responsible to ensure a Lorentz invariant and non-null norm besides carrying much of the physical information encoded on flag-dipole spinors. In such a way, given emergent operator just playing an important role when one deal with phenomenological applications \cite{elkograviton,rodolfoalternative}, cosmological applications and mathematical physics analysis \cite{beyondlounesto,restrictedinomata,juliodual2020,juliodualpla}. Therefore, we highlight its matrix form in addition to exploring some of its main characteristics. 

Notwithstanding, we analyse the fundamental characteristics of the flag-dipole spinors under the action of the $\mathcal{C}$, $\mathcal{P}$ and $\mathcal{T}$ discrete symmetries. Moreover, we conclude that flag-dipole spinors hold $(\mathcal{CPT})^2 = +\mathbbm{1}$, an unexpected behaviour for a spinorial field. However, previously predicted by Wigner in one of his works \cite{Wigner2}, placing the flag-dipole theory in a well-posed physical and mathematical level. Thus, by inspection, we suppose that flag-dipole spinors also belongs to a degenerated Hilbert space. Besides, we leave open windows for an approach like the one used in \cite{elkostates}.

This paper is organized as it follows: In section \ref{capdual} we provide a direct definition of the flag-dipole spinors dual structure, analyzing its main features, besides highlighting verisimilitude with other examples of dual-helicity spinors adjoint structure, present in the current literature. In section \ref{capsimetrias} we advance in the formalism of the $\mathcal{C}$, $\mathcal{P}$ and $\mathcal{T}$ discrete symmetries and also we compute $(\mathcal{CPT})^2$. Finally, in section \ref{remarks} we present some concluding remarks.       

\section{On the flag-dipole dual structure definition}\label{capdual}

Flag-dipole (or type-4) spinors stands for a very rare set of spinors in the literature and which had not been listed in physics applications until recently. They are candidates to construct mass-dimension-one fermions \cite{cavalcanti4} endowed with dual-helicity \cite{chengflagdipole}, and they may explain the reheating phase of the universe \cite{tipo4epjc}.
As soon as spin-$1/2$ mass-dimension-one flag-dipole spinors were explicitly defined in \cite{tipo4epjc}, we now turn our attention to explicitly define the flag-dipole adjoint structure ($\widetilde\Lambda(p^{\mu})$), pointing out the main features encoded on the operator which compose such structure, and evincing some important details that were not previously explained.
  As it can be seen, Dirac's dual structure is well-defined, however, it is not unique and even is not applicable for all spinors. Some cases, such as Elko \cite{jcap} and flag-dipole \cite{tipo4epjc} spinors, for example, require a more involved dual structure. Accordingly, here we provide some details concerning the flag-dipole spinors dual structure. 

If one impose the Dirac's dual structure ($\bar{\psi}=\psi^{\dag}\gamma_0$) to the flag-dipole spinors $(\Lambda(p^{\mu}))$, we face the following norm relation
\begin{eqnarray}
\bar{\Lambda}_{\{\pm,\mp\}}(p^{\mu})\Lambda_{\{\pm,\mp\}}(p^{\mu}) = 0,
\end{eqnarray}
where the lower indexes stands for the right-hand and left-hand component helicity, respectively. Looking towards unveil a hidden physical content, we apply the very same procedure as was previously developed for the Elko spinors in Ref \cite{aaca}.
Thus, this section is reserved for the derivation of a mathematical protocol, and the requirement is a real and invariant norm under Lorentz transformations. 

Let us consider the flag-dipole spinors previously defined in \cite{tipo4epjc}
\begin{equation}\label{flagS}
\Lambda^S_{\{+,-\}}(p^{\mu})=\mathcal{B}_{+}\left(\begin{array}{c}
-\beta^{*-1}_{-}\Theta\phi_L^{-*}(k^{\mu}) \\ 
\beta_+\phi_L^{-}(k^{\mu})
\end{array} \right),
\; \Lambda^S_{\{-,+\}}(p^{\mu})=\mathcal{B}_{-}\left(\begin{array}{c}
\beta^{*-1}_{+}\Theta\phi_L^{+*}(k^{\mu}) \\ 
\beta_-\phi_L^{+}(k^{\mu})
\end{array} \right),
\end{equation} 
and
\begin{equation}\label{flagA}
\Lambda^A_{\{+,-\}}(p^{\mu})=\mathcal{B}_{+}\left(\begin{array}{c}
\beta^{*-1}_{-}\Theta\phi_L^{-*}(k^{\mu}) \\ 
\beta_+\phi_L^{-}(k^{\mu})
\end{array} \right),
\; \Lambda^A_{\{-,+\}}(p^{\mu})=\mathcal{B}_{-}\left(\begin{array}{c}
-\beta^{*-1}_{+}\Theta\phi_L^{+*}(k^{\mu}) \\ 
\beta_-\phi_L^{+}(k^{\mu})
\end{array} \right),
\end{equation} 
where we have defined the Lorentz boost factors as $\mathcal{B}_{\pm} = \sqrt{\frac{E+m}{2m}}\left(1\pm \frac{p}{E+m}\right)$ and the operator $\Theta$ stands for the \emph{Wigner time-reversal operator}, which read \cite{jcap}
\begin{eqnarray}
\Theta = \left(\begin{array}{cc}
0 & -1 \\ 
1 & 0
\end{array}  \right).
\end{eqnarray}
The above spinors satisfy the following orthonormal relations
\begin{eqnarray}
&&\bar{\Lambda}^S_{\{\pm,\mp\}}(p^{\mu})\Lambda^S_{\{\mp,\pm\}}(p^{\mu}) = +2m,\label{ortoS}
\\
&&\bar{\Lambda}^A_{\{\pm,\mp\}}(p^{\mu})\Lambda^A_{\{\mp,\pm\}}(p^{\mu}) = -2m,\label{ortoA}
\\
&&\bar{\Lambda}^S_{\{\pm,\mp\}}(p^{\mu})\Lambda^A_{\{\pm,\mp\}}(p^{\mu}) = 0,\label{orto1}
\\
&&\bar{\Lambda}^S_{\{\pm,\mp\}}(p^{\mu})\Lambda^A_{\{\mp,\pm\}}(p^{\mu}) = 0,\label{orto2}
\\
&&\bar{\Lambda}^A_{\{\pm,\mp\}}(p^{\mu})\Lambda^S_{\{\pm,\mp\}}(p^{\mu}) = 0,\label{orto3}
\\
&&\bar{\Lambda}^A_{\{\pm,\mp\}}(p^{\mu})\Lambda^S_{\{\mp,\pm\}}(p^{\mu}) = 0.\label{orto4}
\end{eqnarray}
The indexes $S$ and $A$ are related with the positive and negative sing on the right-hand side of the norms relations above. Please, note that the orthonormal relations are independent of the phase $(\beta_{\pm})$. However, from now on, we confine ourselves to the only constraint $|\beta_{\pm}|^2 \neq 1$, otherwise, we do not guarantee a proper flag-dipole spinor, in very agreement with \cite{chengflagdipole}. Such a judicious phases constraint lead to a complete set of flag-dipole spinors carrying a non-null and a Lorentz invariant norm. 

%The results displayed in \eqref{ortoS}-\eqref{orto4} suggests, thus, to define the dual spinor as
%\begin{eqnarray}\label{dual1}
%&&\widetilde\Lambda^{S/A}_{\{+,-\}}(p^{\mu}) = \bar{\Lambda}^{S/A}_{\{-,+\}}(p^{\mu}), 
%\nonumber\\
%&&\widetilde\Lambda^{S/A}_{\{-,+\}}(p^{\mu}) = \bar{\Lambda}^{S/A}_{\{+,-\}}(p^{\mu}). 
%\end{eqnarray}
The above relations suggest that the new dual structure must flip the spinor helicity. Looking towards provide a direct definition of the dual structure, the set of relations above make an useful tool to accomplish such task. We start from the very definition of the new dual structure 
\begin{equation}\label{dualflagdipole}
\widetilde\Lambda^{S/A}_{h}(p^{\mu}) \stackrel{def}{=} [\Gamma(p^{\mu})\Lambda^{S/A}_{h}(p^{\mu})]^{\dag}\gamma_0,
\end{equation}
where $h$ stands for the helicity. The $\Gamma(p^{\mu})$ operator must obey a set of requirements. Denoting the spinor space by $\mathcal{S}$, then the $\Gamma(p^{\mu})$ operator is such that 
\begin{eqnarray}
\Gamma(p^{\mu}): \mathcal{S}&\rightarrow& \mathcal{S}, \nonumber\\ \Lambda_h&\mapsto& \Lambda_{h'}. \label{2}
\end{eqnarray} 
Moreover, $\Gamma(p^{\mu})$ has to be idempotent ensuring an invertible mapping. From \eqref{2} we are able to have the following two possibilities: $h=h'$, for which $\Gamma(p^{\mu})=\mathbbm{1}$, has it is the case for the Dirac spinors, or $h\neq h'$ leading to a more involved operator \cite{aaca,vicinity,nondirac,beyondlounesto,juliodual2020}.

With the orthonormal relations \eqref{ortoS}-\eqref{orto4} at hands, one is able to define
\begin{eqnarray}
\Gamma(p^{\mu}) &=& \frac{1}{2m}\big[\Lambda^S_{\{+,-\}}(p^{\mu})\bar{\Lambda}^S_{\{+,-\}}(p^{\mu})+\Lambda^S_{\{-,+\}}(p^{\mu})\bar{\Lambda}^S_{\{-,+\}}(p^{\mu}) 
\nonumber\\
&-&\Lambda^A _{\{+,-\}}(p^{\mu})\bar{\Lambda}^A_{\{+,-\}}(p^{\mu})-\Lambda^A_{\{-,+\}}(p^{\mu})\bar{\Lambda}^A_{\{-,+\}}(p^{\mu})\big],
\end{eqnarray}
in which its matricial form reads
\begin{eqnarray}
\Gamma(p^{\mu}) =\left(\begin{array}{cccc}
-g^{*}(\theta,\beta) & f^*_{1}(\phi,\theta,\beta) & 0 & 0 \\ 
f^{*}_{2}(\phi,\theta,\beta) & g^{*}(\theta,\beta) & 0 & 0 \\ 
0 & 0 & -g(\theta,\beta) & f_2(\phi,\theta,\beta) \\ 
0 & 0 & f_{1}(\phi,\theta,\beta) & g(\theta,\beta)
\end{array}\right).
\end{eqnarray}
Here we have defined the functions $g(\theta,\beta)$, $f_{1}(\phi,\theta,\beta)$ and $f_{2}(\phi,\theta,\beta)$ as follows
\begin{eqnarray}
&&g(\theta,\beta) = \frac{\sin(\theta)}{2}\bigg[\frac{(E+p)\beta^{2}_{+}+(E-p)\beta^{2}_{-}}{m\beta^{}_{+}\beta^{}_{-}}\bigg],\nonumber
\\  
&&f_{1}(\phi,\theta,\beta)=e^{i\phi}\bigg[\frac{(E+p)\cos^2(\theta/2)\beta^{2}_{+}-(E-p)\sin^2(\theta/2)\beta^{2}_{-}}{m\beta_{+}\beta_{-}}\bigg], 
\\
&&f_{2}(\phi,\theta,\beta)=e^{-i\phi}\bigg[\frac{(E-p)\cos^2(\theta/2)\beta^{2}_{-}-(E+p)\sin^2(\theta/2)\beta^{2}_{+}}{m\beta_{+}\beta_{-}}\bigg].\nonumber
\end{eqnarray} 
Note that we fully defined the important operator present in the flag-dipole dual structure. 
The $\Gamma(p^{\mu})$ operator obeys the following requirements: $\Gamma^2(p^{\mu})=\mathbbm{1}$, and the inverse indeed equal itself. In the lights of \cite{beyondlounesto} an useful fact concerning such operator is $[\Gamma(p^{\mu}), \gamma_5]=0$. Remarkably enough, a judicious choice of the phases value, as previously shown in Ref \cite{chengflagdipole}, turn it possible to recover the Elko's $\Xi(p^{\mu})$ operator, in other words, $\Gamma(p^{\mu})\rightarrow \Xi(p^{\mu})$. 

Let us now analyse the behaviour of $\Gamma(p^\mu)$ under Lorentz transformations. For a Lorentz boost we have that
\begin{equation}
 \Lambda_{h}(p^{\mu})=\exp(i\boldsymbol{\kappa\varphi})\Lambda_{h}(k^{\mu}),   
\end{equation}
 with $\boldsymbol{\kappa}^{\dag}=-\boldsymbol{\kappa}$, and then
\begin{equation}\label{invxi}
\left[\Gamma(p^{\mu})\Lambda_{h}(p^{\mu})\right]^\dagger\gamma_0\Lambda_{h}(p^{\mu}) = \Lambda^{\dag}_{h}(k^{\mu})e^{i\boldsymbol{\kappa\varphi}}\Gamma^{\dag}(p^{\mu})\gamma_0 e^{-i\boldsymbol{\kappa\varphi}} \Lambda_{h}(k^{\mu}).
\end{equation}
The expression above provides the following relation
\begin{equation}
\Gamma(p^{\mu}) = e^{i\boldsymbol{\kappa\varphi}}\Gamma(k^{\mu})e^{-i\boldsymbol{\kappa\varphi}}.
\end{equation}
The action of the $\Gamma(p^{\mu})$ operator on the flag-dipole spinors provide the following relations
\begin{eqnarray}
\Gamma(p^{\mu})\Lambda^S_{\{+,-\}}(p^{\mu}) = \Lambda^S_{\{-,+\}}(p^{\mu}),
\\
\Gamma(p^{\mu})\Lambda^S_{\{-,+\}}(p^{\mu}) = \Lambda^S_{\{+,-\}}(p^{\mu}),
\\
\Gamma(p^{\mu})\Lambda^A_{\{+,-\}}(p^{\mu}) = \Lambda^A_{\{-,+\}}(p^{\mu}),
\\
\Gamma(p^{\mu})\Lambda^A_{\{-,+\}}(p^{\mu}) = \Lambda^A_{\{+,-\}}(p^{\mu}).
\end{eqnarray}
Such a direct definition of the dual structure provide additional support to the flag-dipole dual structure previously found \cite{tipo4epjc}. It is worth mentioning that the above approach is important because, first --- in parallel with the Elko's case --- it shows the protocol of how the dual structure for dual-helicity spinors emerges. Consequently, we show some properties of the $\Gamma(p^{\mu})$ operator and, finally, we evinced its explicit form, which is necessary for carrying out studies as the one developed in \cite{elkograviton,rodolfoalternative,beyondlounesto,restrictedinomata}. 

We remark that the prescription contained here clearly leads to a theory where the dual structure presented in \eqref{dualflagdipole} provides a spin sum which contain a term which do not manifest covariantly via Lorentz transformations, consequently, at a quantum field level it brings a non-local quantum field. However, in \cite{tipo4epjc}, and supported by \cite{aaca}, a redefinition in the dual structure bring to light a local theory.

\section{On the $\mathcal{C}$, $\mathcal{P}$ and $\mathcal{T}$ discrete Symmetries: The dual helicity spinors $(\mathcal{CPT})^2$ unconventional outcome.}\label{capsimetrias}
The present section carry some fundamental aspects concerning discrete symmetries and flag-dipole spinors. Bearing in mind that through the relations that will be established here, we are able to connect, at the second quantization level, the junction among dynamics, quantum field and locality structure. In addition, such analysis is extremely relevant when it is intended to approach the flag-dipole spinors via the formalism developed in Ref \cite[and references therein]{elkostates}.  

We start by analysing the behaviour of the flag-dipole spinors under action of the parity operator, given operator can be defined as $\mathcal{P} = m^{-1}\gamma_{\mu}p^{\mu}$ \cite{speranca}. As it can be seen in the current literature, dual-helicity spinors do not obey the Dirac equation \cite{jcap,interplay}. In order to illustrate the procedure, we choose the $\Lambda^{S}_{\{+,-\}}(p^{\mu})$ spinor, and then apply the operator $\gamma_{\mu}p^{\mu}$ on it, to obtain 
\begin{eqnarray}\label{diraconflag1}
m^{-1}\gamma_{\mu}p^{\mu}\Lambda^{S}_{\{+,-\}}(p^{\mu}) &=& m^{-1}[E\gamma_0+\gamma_{j}p^j]\Lambda^{S}_{\{+,-\}}(p^{\mu}), 
\nonumber\\
&=& m^{-1}\Big[E\left(\begin{array}{cc}
0 & 1 \\ 
1 & 0
\end{array} \right) + p \left(\begin{array}{cc}
0 & \boldsymbol{\sigma}\cdot\boldsymbol{\hat{p}} \\ 
-\boldsymbol{\sigma}\cdot\boldsymbol{\hat{p}} & 0
\end{array} \right)\Big]\mathcal{B}_{+}\left(\begin{array}{c}
-\beta^{*-1}_{-}\Theta\phi_L^{-*}(k^{\mu}) \\ 
\beta_+\phi_L^{-}(k^{\mu})
\end{array} \right),
\end{eqnarray}
where $\boldsymbol{\sigma}\cdot\boldsymbol{\hat{p}}$ stands for the helicity operator, previously defined in \cite{tipo4epjc}. When this operator acts on the spinor's components, we get
\begin{eqnarray}
&&\boldsymbol{\sigma}\cdot\boldsymbol{\hat{p}}\;\phi_L^{\pm}(k^{\mu}) = \pm \phi_{L}^{\pm}(k^{\mu}),
\\
&&\boldsymbol{\sigma}\cdot\boldsymbol{\hat{p}}\;[\Theta\phi_L^{\pm *}(k^{\mu})] = \mp \Theta\phi_L^{\pm *}(k^{\mu}).  
\end{eqnarray}
After simple mathematical manipulations, one obtain the following relation
\begin{eqnarray}
m^{-1}\gamma_{\mu}p^{\mu}\Lambda^{S}_{\{+,-\}}(p^{\mu}) = m^{-1}\mathcal{B}_{+}(E-p)\left(\begin{array}{c}
\beta_{+}\phi_L^{-} \\ 
-\beta_{-}^{*-1}\Theta\phi_L^{-*}
\end{array} \right),
\end{eqnarray}
taking into account the Einstein's dispersion relation, we are able to write 
\begin{eqnarray}
\mathcal{B}_{+}(E- p) \rightarrow m\mathcal{B}_{-},
\\
\mathcal{B}_{-}(E+p) \rightarrow m\mathcal{B}_{+},
\end{eqnarray}
and also the following relation among the components
\begin{equation}
\Theta\phi_{L}^{-*} = -\phi_L^{+}, \quad \phi_{L}^{-} = \Theta\phi_{L}^{+*}.
\end{equation}
Then, from eq. \eqref{diraconflag1}, we get
\begin{eqnarray}\label{diracflagfinal}
m^{-1}\gamma_{\mu}p^{\mu}\Lambda^{S}_{\{+,-\}}(p^{\mu}) = \mathcal{B}_{-}\left(\begin{array}{c}
\beta_{+}\Theta\phi_{L}^{+*} \\ 
\beta_{-}^{*-1}\phi_L^{+}
\end{array} \right).
\end{eqnarray}
Note that the flag-dipole spinors do not form a set of eigenspinors of the Dirac operator. In other words, the flag-dipole spinors do not satisfy the Dirac equation. This fact had already been observed previously in \cite{interplay}. Repeating the very same procedure described above, \emph{i.e.}, acting with $\gamma_{\mu}p^{\mu}$ on the r.h.s of the equation \eqref{diracflagfinal}, it leads to
\begin{eqnarray}
m^{-1}\gamma_{\mu}p^{\mu}\mathcal{B}_{-}\left(\begin{array}{c}
\beta_{+}\Theta\phi_{L}^{+*} \\ 
\beta_{-}^{*-1}\phi_L^{+}
\end{array} \right) &=& \mathcal{B}_{+}\left(\begin{array}{c}
-\beta^{*-1}_{-}\Theta\phi_L^{-*}(k^{\mu}) \\ 
\beta_+\phi_L^{-}(k^{\mu})
\end{array} \right),
\nonumber\\
 &=& \Lambda^{S}_{\{+,-\}}(p^{\mu}).
\end{eqnarray}
The above results lead us straightforwardly to conclude that $\mathcal{P}^2$ obey the following relation
\begin{equation}
\mathcal{P}^2 = \mathbbm{1}.
\end{equation}

%%%%%%%%%%%%%%%%%  HERE FINISH THE RED COLOR COMMAND %%%%%%%%%%%%%%%%%%%
%
Now, we focus on the charge-conjugation operator, which can be written as $\mathcal{C} = \gamma_2\mathcal{K}$, where $\mathcal{K}$ stands for the algebraic complex conjugation operation \cite{jcap}. As stated in \cite{tipo4epjc,interplay} flag-dipole spinors do not necessarily hold conjugacy under $\mathcal{C}$. Thus, here we provide a quick 
derivation of such observations:
\begin{eqnarray}
\mathcal{C}\Lambda^{S}_{\{+,-\}}(p^{\mu}) &=& \left(\begin{array}{cc}
0 & i\Theta \\ 
-i\Theta & 0
\end{array} \right)\mathcal{B}_{+}\left(\begin{array}{c}
-\beta^{-1}_{-}\Theta\phi_L^{-}(k^{\mu}) \\ 
\beta^{*}_+\phi_L^{*-}(k^{\mu})
\end{array} \right) 
\nonumber\\
&=& \mathcal{B}_{+}\left(\begin{array}{c}\label{flagconjugado}
i\beta^{*}_{+}\Theta\phi_L^{-*}(k^{\mu}) \\ 
-i\beta_{-}^{-1}\phi_L^{-}(k^{\mu})
\end{array} \right).
\end{eqnarray}
Note that the resulting spinor does not belong to the set of flag-dipoles in \eqref{flagS} and \eqref{flagA}, which reinforces the argumentation in \cite{rodolfohierarchy}. Nonetheless, acting twice with $\mathcal{C}$ operator on equation \eqref{flagconjugado}, provides the following result
\begin{equation}
\mathcal{C}^2 = \mathbbm{1}.
\end{equation}

Finally, the last discrete symmetry concern to the time-reversal operator $\mathcal{T} = i\gamma_5\mathcal{C}$, an anti-unitary operator. With the previous results at hands, the action of such an operator on a flag-dipole spinor is given by
\begin{eqnarray}
\mathcal{T}\Lambda^{S}_{\{+,-\}}(p^{\mu}) &=& i\gamma_5  \mathcal{B}_{+}\left(\begin{array}{c}
i\beta^{*}_{+}\Theta\phi_L^{-*}(k^{\mu}) \\ 
-i\beta_{-}^{-1}\phi_L^{-}(k^{\mu})
\end{array} \right), \nonumber\\
&=& -\mathcal{B}_{+}\left(\begin{array}{c}
\beta^{*}_{+}\Theta\phi_L^{-*}(k^{\mu}) \\ 
\beta_{-}^{-1}\phi_L^{-}(k^{\mu})
\end{array} \right).
\end{eqnarray}
Note again that the result obtained is not compatible with the spinors shown in \eqref{flagS} and \eqref{flagA}, evincing that the flag-dipole spinors do not stand for a set of eigenspinors of time-reversal operator. However, we have
\begin{eqnarray}
\mathcal{T}^2\Lambda^{S}_{\{+,-\}}(p^{\mu}) &=& -i\gamma_5\mathcal{B}_{+}\left(\begin{array}{c}
\beta^{*}_{+}\Theta\phi_L^{-*}(k^{\mu}) \\ 
\beta_{-}^{-1}\phi_L^{-}(k^{\mu})
\end{array} \right),
\nonumber\\
&=& -\mathcal{B}_{+}\left(\begin{array}{c}
-\beta^{*-1}_{-}\Theta\phi_L^{-*}(k^{\mu}) \\ 
\beta_+\phi_L^{-}(k^{\mu})
\end{array} \right),
\end{eqnarray}
which allows us to conclude that
\begin{equation}
\mathcal{T}^2 = -\mathbbm{1}.
\end{equation}

%%%%%%%%%%%%%%%%% HERE ENDS THE TEXCOLOR COMMAND %%%%%%%%%%%%%%%%%%%
Previous results evades Lee \& Wick Theorem 1 in Ref \cite{leewick}, where it is stated that all the local spin-$1/2$ fields hold the relation $\mathcal{P}^2=\mathcal{T}^2$, and as it can be seen, for the flag-dipole spinors we obtained $\mathcal{P}^2\neq\mathcal{T}^2$. In the meantime, a very interesting and unexpected outcome emerges when we compute $\mathcal{(CPT)}^2$ for the $\Lambda(p^{\mu})$ spinors, using the results above, it yields
\begin{equation}
\mathcal{(CPT)}^2 = +\mathbbm{1}.
\end{equation}
Thus, both Elko and flag-dipole spinors show congruous features. Such result combined with the above calculations evince that flag-dipole spinors also belong to the Wigner Class 3 \cite{Wigner2} and may also belong to the degenerated Hilbert space \cite{elkostates}. Note that for the second time this behaviour is reported for a fermion.  

For completeness, we establish the commutation/anti-commutation among the discrete symmetry operators. Accordingly, acting $\mathcal{C}$ from the left on the equation \eqref{diracflagfinal} it yields
\begin{eqnarray}
\mathcal{C}\mathcal{P}\Lambda^{S}_{\{+,-\}}(p^{\mu}) = \mathcal{B}_{-}\left(\begin{array}{c}
i\beta_{-}^{-1}\Theta\phi_{L}^{*+} \\ 
i\beta_{+}^{*}\phi_{L}^{+}
\end{array} \right).
\end{eqnarray}
On the other hand,
\begin{eqnarray}
\mathcal{P}\mathcal{C}\Lambda^{S}_{\{+,-\}}(p^{\mu}) = \mathcal{B}_{-}\left(\begin{array}{c}
-i\beta_{-}^{-1}\Theta\phi_{L}^{*+} \\ 
-i\beta_{+}^{*}\phi_{L}^{+}
\end{array} \right).
\end{eqnarray}
The above two results leads to anti-commutativity for the $\mathcal{C}$ and $\mathcal{P}$ operators for flag-dipole spinors. Applying the same reasoning for the other flag-dipole spinors, one establishes that $\mathcal{C}$ and $\mathcal{P}$ anti-commute for all $\Lambda^{S}_{\{+,-\}}(p^{\mu})$
\begin{equation}
\{\mathcal{C},\mathcal{P}\} = 0. \label{cp}
\end{equation}
Moreover, one may find the following relations for the other operators:
\begin{eqnarray}
\big[\mathcal{C},\mathcal{T}\big] = 0, \label{tc}
\quad
\big[\mathcal{T},\mathcal{P}\big] = 0. \label{tp}
\end{eqnarray}

The only relation which is consistent with what is expected for fermions is the one given by eq. \eqref{cp}. We remark that the above approach is important at a quantum field level and here we provide some details. In the Weinberg framework \cite{weinberg1}, the analysis is developed for Dirac spinors, thus, the quantum field is defined upon spinors which satisfy Dirac dynamics (eigenspinors of parity operator) playing the role of expansion coefficients and, then, the quantum field hold invariance under parity transformation, being in agreement with all the Lee \& Wick expectations. In \cite{elkostates}, the situation is a slightly different from the previous case, where the analysis takes into account the eigenspinors of the charge conjugation operator (Elko spinors) to define the quantum field. As it can be seen, such spinors hold an extra degree of freedom (helicity) when compared with Dirac spinors, by the aforementioned reason Elko spinors do not satisfy the Dirac dynamics. Thus, this was the first reported case to evade Lee \& Wick expectations, in other words, it can be understood as a fermion with bosonic traces. In such framework, as the authors claims, that given quantum field must belong to a degenerated Hilbert space. Now, the flag-dipole spinors shows to be an interesting case, due to the fact that it does not satisfy any discrete symmetry ($\mathcal{C}$, $\mathcal{P}$ or $\mathcal{T}$), do not satisfy the Dirac dynamics, and also hold an extra degree of freedom, as Elko do, being the second reported case that do not match with Lee \& Wick expectations.

\section{Concluding Remarks}\label{remarks}

As previously mentioned, we remark once again the importance of the dual structure direct definition due to its delicate structure. Since the dual structure for spinors endowed with dual-helicity feature is not a trivial structure --- and strongly contrasts with Dirac's dual structure--- then, it is necessary to define a more rigorous approach, accordingly, we have made a detailed inspection of the flag-dipole spinor dual structure. Hereupon, we gave an additional mathematical support in the definition of such a structure, evincing some details of the operator that composes such a dual structure.
The explicit form, besides the properties of the $\Gamma(p^{\mu})$ operator which is part of the flag-dipole dual structure, is extremely necessary so that we can advance in other branches of research, \emph{e.g.}, particle physics phenomenology, cosmology and mathematical physics.

We shall finalize making allusive comments about the flag-dipole spinors behaviour under action of $(\mathcal{CPT})^2$. As previously mentioned, the unexpected outcome obtained in Sect.\ref{capsimetrias} was predicted by Wigner, and later revisited in \cite{elkostates}. Hence, this is the second case reported in the current literature of an entity that fits Wigner's non-standard classes, being a fermion with unexpected features, carrying similar features to the Elko spinors. 

\section{Acknowledgements}
RJBR thanks CNPq Grant N$^{\circ}$. 155675/2018-4 for the financial support. ARA thanks to CAPES for the partial financial support. 
 
\bibliographystyle{unsrt}
\bibliography{refs}

%unsrt

\end{document}